\documentstyle[aps,preprint,overcite]{revtex}
\draft

\newcommand{\bb}{\begin{eqnarray}}
\newcommand{\ee}{\end{eqnarray}}
\newcommand{\p}{\partial}
\newcommand{\br}{{\bf r}}

\newcommand{\erfc}{\mbox{erfc}}
\newcommand{\trans}{{\em trans~}}
\newcommand{\cis}{{\em cis~}}

\title{Polymer Translocation Induced by Adsorption}
\author{P. J. Park and W. Sung}
\address{Department of Physics, Pohang University of Science and Technology,\\
Pohang 790-784, Korea}

\begin{document}

\maketitle
\begin{abstract}
\indent
We study the translocation of a flexible polymer through a pore in
a membrane induced by its adsorption on \trans side of the membrane.
When temperature $T$ is higher than $T_c$, 
the adsorption-desorption transition temperature, attractive interaction 
between polymer and membrane plays little role in 
affecting polymer conformation, leading to translocation time that
scales as $\tau\sim L^3$ where $L$ is the polymer contour length. 
When $T < T_c$, however, the translocation time undergoes 
a sharp crossover to $\tau\sim L^2$ for sufficiently long polymers,
following the second order conformational (adsorption) transition. 
The translocation time is found to exhibit the crossover around $T=T_c'$,
which is lower than $T_c$ for polymers shorter than a critical length($N<N_c$).
\end{abstract}

\newpage
\section{INTRODUCTION}

  Biopolymers exhibit various transmembrane conformational states 
determined by their intrinsic properties as well as by the 
interactions with the surrounding environments.
For example, polymer flexibility, capability of hydrogen bond formation 
between segments, and hydrophobic interaction with membrane play crucial 
roles in conformation of membrane-bound proteins.
An important aspect of membrane-protein interaction
can be found in the dynamics of proteins. 
As the experimental observations of protein translocation are
available only in limited situations\cite{SD},
theoretical investigations are also in a primitive stage.
However, a few studies show that polymer
translocation across the membrane is significantly affected
by the surrounding environments and membrane-protein 
interactions\cite{Sung,Baumgartner}. 
Theoretical studies have been focused on the driving
mechanisms of polymer translocation across the membrane
such as chemical potential asymmetry\cite{Sung} and its random
modulation\cite{Park}, 
chaperon bindings\cite{Sung,SPO},
and lipid molecular density imbalance\cite{Baumgartner}. 
It was shown that a minute chemical potential bias per segment of the chain
can bring about a significant effect on translocation, due to the 
long-chain connectivity\cite{Sung,Park,Park2}. In this paper, we
investigate the effect of attractive membrane-polymer
interaction yielding adsorption transition as another
possible driving mechanism.

  To highlight the effects arising from long chain nature,
we consider a rigid plane of negligible thickness
with a hole small enough to allow
only a single segment passage, as a minimal model of
membrane with a translocation pore as shown Fig.~1.  
In this model,
polymer segments interact with the membrane 
attractively in the \trans side, while those in the \cis side recognizes the
membrane only as an impenetrable wall. 
No specific interaction between the segments and pore is assumed.
Furthermore we neglect the mutual interaction between the segments
because we are interested in the most dominant effects resulting from 
the chain connectivity.
The attraction,
if it is strong enough to bring about desorption-adsorption transition, 
would drive the polymer to the \trans side as shown
in Fig.~1 to minimize the free energy.

  The total free energy of the transmembrane polymer has
two distinct contributions. One is the entropic contribution
affected by the steric interaction with the membrane, 
and the other is the adsorption energy due to
attraction in \trans side. At high temperatures, thermal
fluctuation dominates and the polymer
conformation is nearly the same as the case without attractive 
interaction. As the polymer is situated in nearly symmetric
environments in this case, it is driven to either side of the
membrane with equal probability. At temperatures 
below the adsorption-desorption transition temperature, on
the other hand, polymer conformation in the \trans side becomes 
collapsed(adsorbed) and 
the polymer is driven spontaneously to the \trans side
to minimize the total free energy. 
Near the transition temperature, however,
competition between thermal fluctuation and membrane attraction
determines the chain conformation, while the translocation dynamics 
is governed by the interplay between thermal diffusion and 
segmental energy bias due to interaction with membrane.

  In section II, conformation of transmembrane polymer and the associated 
free energy are derived as a function of temperature and interaction
potential parameters. In section III, we consider the dynamics of polymer
translocation as a stochastic process crossing the free energy barrier,
where the effects of attraction and temperature on translocation
dynamics are discussed. Section IV
gives summary and conclusion.

\section{STATISTICAL MECHANICS OF A POLYMER
ONE-END-ANCHORED ON ATTRACTIVE SURFACE}

The transmembrane polymer can be regarded as two independent chains
anchored at the spot of pore(Fig.~1).
In this section, we consider the chain conformation and free energy of
one-end-anchored chain to calculate the free energy 
of a transmembrane polymer.
The Green's function of a polymer $G(\br,\br';n)$ is defined as 
the statistical weight of an $n$-segment 
polymer whose end positions are located at
$\br$ and $\br'$ respectively.
For a polymer with Kuhn length $b$,
$G(\br,\br';n)$ satisfies the following differential equation:
\bb
\left[ \frac{\p}{\p n} -\frac{b^2}{6}\nabla^2 \right] G(\br,\br';n)=
\delta(\br-\br').
\label{eq:Edwards} 
\ee
In unbound free space it is reduced to a familiar Gaussian form:
\bb
G_0(\br,\br';n) = \left[\frac{2\pi n b^2}{3}\right]^{-3/2}
\exp\left[- \frac{3(\br-\br')^2}{2 n b^2}\right].
\label{eq:G0}
\ee

For a polymer subjected to an attractive wall($z=0$ surface),
the attractive potential effect can be incorporated 
in terms of the following boundary condition\cite{deGennes},
\bb
\left[ \frac{\p}{\p z} \log G(\br,\br';n)\right]_{z=0}=c.
\label{eq:GBC}
\ee
This boundary condition consideration is valid only for the
case that the attractive interaction range is short enough
compared with other length scales involved such as polymer radius of gyration
or adsorption layer thickness.
The $c$ is a convenient parameter descriptive of surface interaction
the details of which this mesoscopic description usually bypasses.
Although, for flexible chain, the Edwards equation is more adequate
to describe the adsorbed state incorporating details of surface interaction,
this mesoscopic description, also applicable to semiflexible chains,
can easily yield useful results for a full range of temperatures.
For the cases of weak adsorption $u_0\ll k_BT$ and flexible chain $a\gg b$
considered throughout this paper,
we will use the Edwards equation to determine the 
{\it interaction parameter} $c$
in terms of temperature, attraction strength($u_0$)
and range($a$).

The Green's function in half space($z>0$) can be written as
\bb
    G(\br,\br';n) = G_0(x,x';n)G_0(y,y';n)G(z,z';n)
\ee
where $G_0(x,x';n)$ and $G_0(y,y';n)$ are one dimensional Gaussian
propagators analogous to Eq.~\ref{eq:G0}, and the $z$ component is 
given by\cite{BH,LR,EKB}
\bb
&&G(z,z';n) = \left[\frac{\pi nb^2}{6}\right]^{-1/2}
\left[\frac{1}{2}\left\{ \exp\left\{-\frac{3(z-z')^2}{2nb^2}\right\} + \exp\left\{-\frac{3(z+z')^2}{2nb^2}\right\}\right\} \right. \nonumber \\
&&-\pi^{1/2}w 
\exp\left\{ -\frac{3(z+z')^2}{2nb^2} + \frac{3(z+z'+cnb^2/3)^2}{2nb^2} \right\} 
\left. \erfc\left\{ \left(\frac{2nb^2}{3}\right)^{-1/2} (z+z'+cnb^2/3) \right\}\right]
\ee
where $\erfc(u) \equiv (2/\sqrt{\pi}) \int_u^{\infty}\exp(-u'^2)du'$
is the complementary error function, and
\bb
w\equiv (n/6)^{1/2}bc.
\ee
The $z$ component of Green's function has the following limiting behaviors:
\bb
&G(z,z';n)=G_0(z,z';n),& \quad z,z' \rightarrow \infty,\\
&G(z,z';n)=\left[ \frac{2\pi n b^2}{3} \right]^{-1/2} \left[ \exp\left(-\frac{3(z-z')^2}{2nb^2}\right) - \exp\left(-\frac{3(z+z')^2}{2nb^2}\right) \right], & 
\quad w \gg 1,
\ee
as it should be.
The latter is the ideal chain Green's function 
in the presence of absorbing boundary at $z=0$.
Now that $|w|$ is defined as the ratio of free polymer size($\sim n^{1/2}$) 
and adsorbed layer thickness($\sim |c|^{-1}$) for low temperature
adsorbed states($w<0$, $c<0$),
it can be regarded as a parameter 
describing the effects of surface attraction or temperature.
For high temperature desorbed states($w>0$), $w$ measures the strength 
of steric repulsion, which also entails the surface attraction effect.

The conformational partition function of a polymer with
one end anchored on attractive surface is given by\cite{EKB}
\bb
Q(n) &\equiv& \lim_{z'\rightarrow 0} \int_0^\infty G(z,z';n) dz\\
&=& \exp(w^2)\erfc(w) \label{eq:QQ}\\
&=& \left\{ \begin{array}{ll}
		\pi^{-1/2}w^{-1},& w\gg 1\\
		1-2\pi^{-1/2}w,& w\simeq 0\\
		2\exp(w^2),& w\ll -1,
	\end{array} \right.
\ee
where the bulk value of partition
function $q^n$ is omitted throughout this paper, where $q$ is
the partition function of a single segment.
To see the conformational states of anchored polymer explicitly,
let us see the conditional probability distribution of free end defined by
\bb
P_n(z|z'=0)  \equiv\frac{1}{Q(n)} G(z,z'=0;n),
\ee
which has the following limiting forms:
\bb
P_n(z|z'=0)=  \left\{ \begin{array}{ll}
                (nb^2/3)^{-1} z \exp(-3z^2/2nb^2),& w\gg 1\\
                (\pi nb^2/6)^{-1/2}\exp(-3z^2/2nb^2),&  w= 0\\
                |c|\exp(-|c|z),& w\ll -1.
        \end{array} \right.
\ee
For $w\gg 1$, $P_n(z|z'=0)$ is equal to the distribution without attraction,
which corresponds to the high temperature desorbed phase. On
the other hand, for $w \ll -1$, $P_n(z|z'=0)$ represents the
distribution for low temperature adsorbed phase of a polymer
subject to a surface
of sufficiently short range attraction. The distribution at the
adsorption-desorption transition point corresponds to the case $w = 0$.
 
Using the partition function of a one-end-anchored polymer in Eq.~\ref{eq:QQ},
its free energy is given by
\bb
   F(n) &=& -k_B T \log Q(n)  \\
	&=& -k_BT\left[ w^2+\log\erfc(w)\right]\\
	&\simeq& \left\{ \begin{array}{ll}
		(k_BT/2) \log n + \mbox{const},& w\gg1\\
		2k_BTn^{1/2}bc(6\pi)^{-1/2}, & w\simeq 0\\
		-k_BT nb^2c^2/6+ \mbox{const},& w\ll -1.
	\end{array} \right.
\ee
Note that the $F(n)$ changes sign at $w=0$, which corresponds to
adsorption-desorption transition point.
At high temperature($w\gg 1$), we can see the 
polymer free energy, the value relative to that of free polymer,
goes like $\log n$ due to the anchorage effect.
At low temperatures($w\ll -1$), adsorption free energy scales as $n$
involving segmental free energy as $-k_BT b^2c^2/6$.

  Since the transmembrane polymer is composed of two independent 
one-end-anchored chains with length $n$ and $N-n$ in
\cis and \trans side respectively(Fig.~1), its partition function
is given by
\bb
        {\cal Q}(n) \equiv Q_{cis}(N-n)Q_{trans}(n,c),
\ee
where the effect of attraction appears only in the \trans side
partition function in terms of $c$.  This partition
function has the following limiting behaviors:
\bb
{\cal Q}(n) \sim \left\{ \begin{array}{ll}
		(N-n)^{-1/2}n^{-1/2},&T\gg T_c\\
		(N-n)^{-1/2},&T=T_c\\
		(N-n)^{-1/2}\exp(nb^2c^2/6),&T\ll T_c.
	\end{array} \right.
\ee
The free energy as a function of {\it translocation coordinate} 
$n$, defined as the number of segments in \trans side, is,
apart from additive constants, 
\bb
{\cal F}(n)&=& -k_BT \log {\cal Q}(n)\\
&=&\frac{1}{2}k_BT\log(N-n) -k_BT \left[\frac{nb^2c^2}{6}+\log\erfc(\sqrt{\frac{n}{6}}bc)\right] \label{eq:FofC}\\
&=& \left\{ \begin{array}{ll}
		\frac{1}{2}k_B T \log\left[(N-n)n\right],&T\gg T_c\\
		\frac{1}{2}k_BT\log(N-n) ,&T=T_c\\
		\frac{1}{2}k_BT\log(N-n)-k_BT\frac{nb^2c^2}{6},&T\ll T_c.
	\end{array} \right.
\ee
In Fig.~2, this free energy function is depicted for different 
values of $c$. ${\cal F}(n)$ exhibits nearly symmetric barrier for
positive $c$($T>T_c$), while it does exhibits a linearly slant shape for
negative $c$($T<T_c$). This signifies that the polymer is driven to
the \trans side due to the adsorption for $c < 0$. One remarkable 
point here is that ${\cal F}(n)$ slants to the \trans side at
$T = T_c$ even though the polymer in the \trans side is not
in its adsorbed state. The $n$ segments in the \trans side
feel neither steric interaction nor attraction. The cancelation 
of these two effects makes the total free energy 
be determined only in terms of the $N-n$ segments in the \cis side.\\

\noindent \underline{{\bf Determination of c}}

  In the previous section, the free energy of a transmembrane polymer is
determined in terms of interaction parameter $c$, which is 
defined as the logarithmic derivative of polymer Green's function 
at the surface(Eq.~\ref{eq:GBC}). 
To determine $c$ in its adsorbed phase in terms of interaction potential 
parameters, we consider a microscopic model of flexible polymer 
weak adsorption in this
subsection. As the polymer conformation in its adsorbed phase
is not affected by the anchorage,
we will consider the adsorbed polymer without anchorage for simplicity.

Let us choose the interaction potential between a monomer and the membrane as 
\bb
          v(\br) =  \left\{ \begin{array}{cc}
			 -u_0, & 0<z<a,\\
			0, & z>0,
		\end{array} \right.
\ee
where $u_0 > 0$ is the strength and $a$ is the range of attraction. 
Under the conditions $u_0 \ll k_BT$ and $a \gg b$\cite{Kuznesov}, 
the polymer Green's function obey the Edwards equation
\bb
\left[ -\frac{b^2}{6}\frac{\p^2}{\p z^2} + \beta v(z) \right] G(z,z';n)
= -\frac{\p}{\p n} G(z,z';n).
\ee
It can be expanded as
\bb
     G(z,z';n) = \sum_i  \exp (-n\lambda_i) \psi_i(z) \psi_i(z' ) 
\ee
where $\lambda_i$ and $\psi_i(z)$ are the eigenvalues and eigenfunctions
of the Schr\"{o}dinger-like equation,
\bb
\left[ -\frac{b^2}{6} \frac{\p^2}{\p z^2} + \beta v(z)\right] \psi_i(z) 
= \lambda_i\psi_i(z),
\label{eq:eigen}
\ee
subject to the boundary condition
\bb
               \psi_i(z = 0) = 0.
\ee
By  solving  this  eigenvalue  problem,  we  can  get  the
adsorption-desorption transition temperature $T_c$ and the
value of $c$ as a function of $T$, $a$ and $u_0$ for adsorbed state.

  For $T < T_c$, the ground state dominance approximation,
$G(z,z';n) \simeq  \exp (-n\lambda_0) \psi_0(z) \psi_0(z' )$,
works well for a long chain, 
where $\lambda_0$ is the lowest eigenvalue and  $\psi_0(z)$ is the 
corresponding eigenfunction of Eq.~\ref{eq:eigen}. 
The ground state eigenfunction in $z$ direction is given by
\bb
 \psi_0(z) = \left\{ \begin{array}{ll}
                         A\sin(kz), & 0<z<a,\\
                         B\exp(-\kappa z), & z>a,
                \end{array} \right.
\ee
where $k =  (6/b^2)^{1/2} (\beta u_0 + \lambda_0 )^{1/2}$
and  $\kappa = (6/b^2)^{1/2} |\lambda_0|^{1/2}$   with
$\lambda_0<0$ for $T < T_c$. 
The transition temperature $T_c$ can be determined as
\bb
k_BT_c=\frac{24u_0a^2}{\pi^2b^2},
\label{eq:Tc}
\ee
below which a bound state of negative eigenvalue exists.
The continuity condition of $\psi_0(z)$ and
$\psi_0'(z)$ at $z =a$ gives the following relation:
\bb
\kappa a = -ka \cot(ka)
\label{eq:cont}
\ee
which determines $\lambda_0$ for $T<T_c$. 
The interaction parameter $c$ for $T<T_c$ can be
defined as
\bb
c \equiv \left[ \frac{1}{\psi_0(z)} \frac{\p}{\p z} \psi_0(z) \right]_{z=a}
= -\kappa.
\label{eq:c}
\ee
This microscopic definition of $c$ can be identified with the previous
definition if the interaction is short-ranged.
From the solution of Eq.~\ref{eq:cont} and Eq.~\ref{eq:c}, 
$c$ can be written as the following functional form for $T\le T_c$:
\bb
c = -a^{-1} H(\alpha)
\label{eq:scaling}
\ee
where $\alpha \equiv (a/b) (6\beta u_0)^{1/2}$, and $H(\alpha)$
is depicted in Fig.~3.
As $T$ approaches $T_c$ from below, $\alpha$ goes to $\pi/2$ and $H(\alpha)$
goes to zero, which yields
\bb
c\simeq -\frac{\pi^2}{8a} \frac{T_c-T}{T_c} + {\cal O}(T_c-T)^2,
\label{eq:cNearTc}
\ee
for $T\lesssim T_c$.
At sufficiently low temperatures($T\ll T_c$), on the other hand,
$\alpha \gg 1$ and $H(\alpha) \simeq \alpha$.
In this regime, $c$ can be approximated as
\bb
c \simeq -b^{-1}(6\beta u_0)^{1/2} \sim -T^{-1/2},
\label{eq:cLowT}
\ee
which is independent of interaction range $a$.
Eq.~\ref{eq:cNearTc} and Eq.~\ref{eq:cLowT} are useful to examine
the temperature dependence of polymer descriptions
derived from the boundary condition(Eq.~\ref{eq:GBC}), which
will be demonstrated in the next section.

\section{DYNAMICS OF POLYMER TRANSLOCATION}

  As shown by us\cite{Sung}, the translocation of a polymer can be
thought of as a one dimensional diffusive process of the
translocation coordinate $n$ crossing the effective potential 
${\cal F}(n)$.  The probability density of $n(t)$, $P(n,t)$, is described by
Fokker-Planck equation
\bb
\frac{\p}{\p t} P(n,t) = {\cal L}_{FP}(n) P(n,t),
\ee
where ${\cal L}_{FP}(n)$ is the Fokker-Planck operator given by
\bb
{\cal L}_{FP}(n)\equiv \frac{1}{b^2}\frac{\p}{\p n}D(n)\exp[-\beta {\cal F}(n)]
\frac{\p}{\p n} \exp[\beta {\cal F}(n)],
\ee
with $D(n)$ defined as the whole chain diffusion coefficient. The translocation 
time of a polymer can be defined as the mean first passage time
$\tau(n; n_0)$, the time for diffusion from $n_0$ to $n$,
which satisfies\cite{SSS81}
\bb
{\cal L}_{FP}^{\dag}(n_0) \tau(n; n_0) = -1,
\label{eq:backward}
\ee
where ${\cal L}_{FP}^{\dag}(n_0)$ is the backward Fokker-Planck operator
given by
\bb
{\cal L}_{FP}^{\dag}(n_0)\equiv \frac{1}{b^2}
\exp[\beta {\cal F}(n_0)] \frac{\p}{\p n_0}D(n_0)
\exp[-\beta {\cal F}(n_0)] \frac{\p}{\p n_0}. 
\ee
Using the following boundary conditions
\bb
\frac{\p}{\p n_0} \tau(n; n_0=1)=0\\
\tau(n; n_0=N-1)=0,
\ee
the solution of the above backward equation(Eq.~\ref{eq:backward}) 
can be obtained and the polymer translocation time is determined as
\bb
\tau &\equiv &\tau(N-1;1)\\
&=& b^2\int_1^{N-1} dn \frac{1}{D(n)} e^{ \beta {\cal F}(n) }
\int_1^{n} dn' e^{ -\beta {\cal F}(n') }.
\label{MFPT}
\ee
By putting the diffusion coefficient $D(n)$ 
to be a constant $D$ for simplicity, 
the translocation time as a function of $w$ is shown in Fig.~4,
where a drastic change in $\tau$ occurs near $w = 0$.  Because $w$ or $c$ is
expected to be a monotonic function of temperature 
as shown in Eq.~\ref{eq:scaling} and Fig.~3, Fig.~4 can be understood 
as a sharp dynamic crossover induced by temperature change.
Note that $\tau \sim |w|^{-2} \sim |c|^{-2}$ for $T\ll T_c$, 
while $\tau$ changes smoothly only by a small amount for $T>T_c$.

The Fig.~5 is a plot of $\tau$, which is obtained 
using the free energy(Eq.~\ref{eq:FofC}) 
with $c$ given by Eq.~\ref{eq:scaling}, 
as a function of $N$ for four different temperatures
ranging from $T\gg T_c$ to $T\ll T_c$.
For $T \gg T_c$, i.e., $\beta u_0 \simeq 0$, the transmembrane polymer
does not feel attractive interaction and translocation time(Eq.~\ref{MFPT})
is reduced to
\bb
\tau \simeq {\pi^2 \over 8} \left[ \frac{L^2}{2D} \right] \sim L^{3},
\ee
where $L = Nb$ is the contour length of the polymer and the
diffusion coefficient is given by $D = k_BT/(N\gamma)$, with segmental
hydrodynamic friction coefficient $\gamma$ following
the Rouse model\cite{Doi}. In this high temperature regime,
translocation time is proportional to $L^3$(Fig.~5, A),
which is the same result as the case
without the attractive interaction\cite{Sung}.

  As the temperature is lowered, the free energy curve tends to slant
down to the right and the prefactor of $\tau$
decreases smoothly until $T = T_c$ as shown in Fig.~4. 
At $T = T_c$, the translocation 
time becomes
\bb
\tau = {2 \over 3} \left[ \frac{L^2}{2D} \right] \sim L^{3},
\ee
where the prefactor is reduced from $\pi^2/8$ to $2/3$. 
In this case, the steric interaction in the \trans side is
canceled off by the attractive interaction, which drives the polymer
to the \trans side, but $\tau$ is still proportional to $L^3$(Fig.~5, B).
Assuming $D$ increases as $T$ increases like $D\sim T^\alpha$ with $\alpha>0$,
\bb
\frac{\tau(T\gg T_c)}{\tau(T=T_c)} \simeq \frac{3\pi^2}{16} \left(\frac{T_c}{T}\right)^\alpha \ll 1.
\ee
Thus, as can be seen in A and B of Fig.~5,
the $\tau$ increases as $T$ decreases to $T_c$,
which results from the suppression of thermal diffusion.
The $\tau$ attains its maximum at $T=T_c$,
since for $T<T_c$, the polymer adsorption speeds the translocation
as discussed below.

  For $T \ll T_c$, $c$ behaves as $c\sim -T^{-1/2}$(Eq.~\ref{eq:cLowT}) 
and the translocation time(Eq.~\ref{MFPT}) is given by
\bb
\tau &\simeq& \left[\frac{L^2}{2D}\right] \frac{2}{N b^2 c^2/6}\\
&=& \left[\frac{L^2}{2D}\right] \frac{2}{N\beta u_0}\\
&=& N^2b^2\gamma/u_0 \label{eq:plateau}\\
&\sim& L^2,
\ee
which is depicted in Fig.~5, D.
In this low temperature regime, the free energy of transmembrane polymer 
is dominated by the attraction energy which is linearly proportional to $n$,
where the prefactor of $n$ can be interpreted as the energy per segment 
in \trans side, or transmembrane segmental energy bias that features as
``chemical potential difference" between the two sides of membrane\cite{Sung}.

We find that the crossover of scaling behavior 
from $\tau\sim L^3$ to $\tau\sim L^2$ 
sharply occurs around $T=T_c$,
which can also be inferred from Fig.~4.
The conditions for the crossover can be analyzed in detail as follows
using the Edwards equation approach for $T<T_c$.
As shown by us\cite{Sung}, segmental energy bias approximately
larger than $k_BT/N$ can bring about the crossover in translocation
from $\tau\sim L^3$ to $\tau\sim L^2$
due to chain connectivity(cooperativity).
For the adsorption-driven translocation when $T<T_c$, 
the condition for the crossover can be written as
\bb
|\lambda_0| \gtrsim \frac{1}{N},
\label{eq:condition}
\ee
where $\lambda_0$, the lowest eigenvalue in Eq.~\ref{eq:eigen}, 
is the segmental free energy in units of $k_BT$ in adsorbed state.
This condition(Eq.~\ref{eq:condition}) for $T$ near $T_c$ 
can be converted to
\bb
T \lesssim T_c'(N) =
T_c\left[1-\frac{8a}{\pi^2} \left(\frac{6}{Nb^2} \right)^{1/2} \right],
\ee
where the relation $\lambda_0 = -b^2H^2(\alpha)/6a^2$(see Eq.~\ref{eq:scaling})
was used.
This condition implies that the crossover occurs at 
$T_c'$ smaller than $T_c$ for finite chain length $N$.
That is, for $T_c' <T<T_c$, $\tau\sim L^3$
although the transmembrane polymer is adsorbed in \trans side. 
For a numerical example for the difference between $T_c$ and $T_c'$, 
we consider the case, $a/b=10$ and $N=60000$, 
which yields $T_c' \approx 0.9T_c$.
The C of Fig.~5 depicts this situation.
For temperatures less than $T_c$,
the condition of Eq.~\ref{eq:condition} yields the critical chain length
\bb
N_c(T) \simeq \frac{384a^2}{\pi^4b^2} \left(\frac{T_c-T}{T_c}\right)^{-2} 
\ee
around which the crossover from $\tau\sim L^3$ to $\tau\sim L^2$ occurs.
Indeed this $N_c$ is consistent with the crossover point in C of Fig.~5.

Table 1 summarizes the conditions for the crossover
and associated chain conformations for diverse temperatures 
and chain length.
The meaning of
two critical temperatures $T_c$, $T_c'$ as well as the critical chain length
$N_c$ for chain cooperativity, can be understood as follows.
The segmental free energy $\lambda_0(T)$ vanishes at $T=T_c$,
while $N |\lambda_0(T)|$ becomes larger than $1$ for $T \lesssim T_c'$ 
or $N\gtrsim N_c$, 
which are the conditions for a cooperative, facilitated translocation
driven by an energetic bias.
In this study, we do not incorporate the solvent effect 
which will induce interaction between segments and thus affect
the polymer free energy function significantly.
Yet the crossover in $\tau$ is seen to occur irrespective of excluded volume
effect, as a consequence of the long-chain nature of the polymer 
interacting with the membrane\cite{Park2}. Incorporation of the effect will
modify only $T_c$ and $T_c'$.

\section{SUMMARY AND CONCLUSION}

  We considered the dynamics of flexible polymer translocation
as a diffusive process crossing the free energy barrier
which we derived in section II.
The translocation dynamics is found to be strongly correlated with 
transmembrane chain conformation, both of which are varied depending upon
the competition between thermal fluctuation and 
attractive polymer interaction with the membrane.
The polymer adsorption-desorption transition point($T_c$)
divides the length scaling regimes of translocation time,
$\tau\sim L^3$ and $\tau\sim L^2$, for sufficiently long chains($N>N_c$).
However, for shorter chains($N<N_c$), a crossover of two regimes
is found to occur at a lower temperature, $T_c'$, 
although the conformational transition still occurs 
at the same transition temperature, $T_c$.
We used the mesoscopic theory of polymer adsorption using the machinery of 
boundary condition, supplemented by the microscopic Edwards equation approach
to consider the case of flexible chain weak adsorption in detail.
This theory can be adapted to the more biologically-relevant situations
of semiflexible chain strong adsorption,
if the interaction parameter can be appropriately given.

\section{Acknowledgments}
We acknowledge the support from POSTECH BSRI special fund,
Korea Science Foundation(K96004), and BSRI program(N96093), Ministry of
Education.


\newpage

\begin{figure}[h]
\caption{Translocation of a polymer through a pore. Polymer segments are
adsorbed in the \trans side due to the attractive interaction at sufficiently
low temperatures.}
\end{figure}

\begin{figure}[h]
\caption{Free energy ${\cal F}(n)$ as a function of 
translocation coordinate $n$.
(A. $bc=1.0$, B. $bc=0$, C. $bc=-0.5$, D. $bc=-1.0$)}
\end{figure}

\begin{figure}[h]
\caption{$H(\alpha)$ versus $\alpha\equiv (a/b)(6\beta u_0)^{1/2}$.
At $T=T_c$($\alpha=\pi/2$), $H(\alpha)=0$ and $c\equiv -a^{-1}H(\alpha)=0$.
For $T \ll T_c$($\alpha\gg 1$), $H(\alpha) \simeq \alpha$
and $c \simeq -b^{-1}(6\beta u_0)^{1/2}$(dotted line).}
\end{figure}

\begin{figure}[h]
\caption{Translocation time $\tau$ normalized by $\tau_0 \equiv L^2/(2D)$
versus $|w|$, with $w\equiv bc(N/6)^{1/2}$, 
(A. $c>0(T>T_c)$, B. $c<0(T<T_c)$).
Note that $\tau/\tau_0 \sim w^{-2}$ for $T\ll T_c$.
Inset: The same figure in linear scale, 
where sharp transitional behavior occurs
at $c=0(T=T_c)$.}
\end{figure}

\begin{figure}[h]
\caption{Translocation time $\tau$ in units of $\tau_1 \equiv \gamma b^2/u_0$
versus chain length $N$, where $a=10b \ll Nb$, 
(A. $T=10T_c$, B. $T=T_c$, C. $T=0.9T_c$, D. $T\ll T_c$).
$\tau\sim L^3$ for A and B, and $\tau\sim L^2$ for D.
Crossover between them can be seen for C near $N=N_c$
as the chain length varies.}
\end{figure}

\newpage
\noindent Table 1. Length scaling behavior of polymer translocation time $\tau$
and associated chain conformations
as a function of temperature and chain length.\\
\begin{center}
\begin{tabular}{||c|c||c|c||}
\hline \hline 
      &    ~~shorter chain($N<N_c$)~~  &  & ~~longer chain($N>N_c$)~~ \\
\hline 
$T>T_c$ & $\tau \sim L^3$ (desorbed) & $~~T>T_c~~$ & $\tau \sim L^3$ (desorbed)\\
\hline
~~$T_c'<T<T_c$~~ & $\tau \sim L^3$ (adsorbed)& $T<T_c$&$\tau\sim L^2$ (adsorbed)\\
\hline
$T<T_c'$ & $\tau\sim L^2$ (adsorbed) &    &     \\
\hline
\hline
\end{tabular}
\end{center}

\end{document}